\documentclass[prc,amsmath,twocolumn,showpacs,superscriptaddress]{revtex4}
\usepackage[dvips]{graphicx}
\usepackage{dcolumn}
\usepackage{comment}
\usepackage{amssymb}

\newcommand{\bea}{\begin{eqnarray}}
\newcommand{\eea}{\end{eqnarray}}
\newcommand{\be}{\begin{equation}}
\newcommand{\ee}{\end{equation}}
\newcommand{\eq}[1]{Eq.~(\ref{#1})}
\newcommand{\bwt}{\begin{widetext}}
\newcommand{\ewt}{\end{widetext}}

\renewcommand{\vec}[1]{\boldsymbol{#1}}
\newcommand{\im}{{\mathrm{i}}}

\newcommand{\expl}[1]{{\mathrm{exp}}\left[#1\right]}
\renewcommand{\exp}[1]{{\mathrm{e}}^{#1}}
\newcommand{\dif}{{\mathrm{d}}}

\newcommand{\dderiv}[1]{\frac{\dif^2}{\dif #1^2}}

\newcommand{\bra}[1]{\langle#1|}
\newcommand{\ket}[1]{|#1\rangle}
\newcommand{\braket}[2]{\langle#1|#2\rangle}
\newcommand{\matrixel}[3]{\langle#1|#2|#3\rangle}
\newcommand{\reduced}[3]{\langle#1 \parallel #2 \parallel #3\rangle}

\newcommand{\conkvc}{\frac{2\mu_{vc}}{\hbar^2}}
\newcommand{\convc}{\frac{\hbar^2}{2\mu_{vc}}}
\newcommand{\conpt}{\frac{\hbar^2}{2\mu_{Pt}}}

\newcommand{\TR}{T_{\vec{R}}}
\newcommand{\Tr}{T_{\vec{r}}}
\newcommand{\coords}{\vec{R},\vec{r},\xi}

\newcommand{\cg}[6]{(#1 #2 #3 #4 | #5 #6 )}

\newcommand{\threej}[6]{\begin{pmatrix}#1&#2&#3\\#4&#5&#6\end{pmatrix}}
\newcommand{\threejz}[3]{\begin{pmatrix}#1&#2&#3\\0&0&0\end{pmatrix}}
\newcommand{\sixj}[6]{\begin{Bmatrix}#1&#2&#3\\#4&#5&#6\end{Bmatrix}}
\newcommand{\ninej}[9]{\begin{Bmatrix}#1&#2&#3\\#4&#5&#6\\#7&#8&#9\end{Bmatrix}}

\newcommand{\tripolar}{( C_{\Lambda}(\vec{\hat{R}}) \cdot \{ C_{\Lambda'}(\vec{\hat{r}})
\otimes C_{Q}(\xi) \}_{\Lambda} )}

\newcommand{\fig}[1]{Fig.~\ref{#1}}

\newcommand{\yes}{\checkmark}
\newcommand{\no}{$\times$}
\newcommand{\formp}{P_{a:a'}^{KQ\lambda:\Lambda}}
\newcommand{\formf}{F_{J_Pin:J_P'i'n'}^{\Lambda}}

\begin{document}

\title{XCDCC: Core Excitation in the Breakup of Exotic Nuclei}

\author{N.~C.~Summers}
 \affiliation{National Superconducting Cyclotron Laboratory,
Michigan State University, East Lansing, Michigan 48824, U.S.A.}
\author{F.~M.~Nunes}
 \email{nunes@nscl.msu.edu}
 \affiliation{National Superconducting Cyclotron Laboratory,
Michigan State University, East Lansing, Michigan 48824, U.S.A.}
 \affiliation{Department of Physics and Astronomy,
Michigan State University, East Lansing, Michigan 48824, U.S.A.}
\author{I.~J.~Thompson}
 \affiliation{Department of Physics, University of Surrey,
 Guildford, GU2 5XH, U.K.}
\date{\today}

\begin{abstract}
The eXtended Continuum Discretized Coupled Channel (XCDCC) method is developed
to treat reactions where core degrees of freedom play a role.
The projectile is treated as a multi-configuration coupled channels
system generated from a valence particle coupled to a deformed core which is allowed to excite.
The coupled channels initial state breaks up into
a coupled channels continuum which
is discretized into bins, similarly to the original CDCC method.
Core collective degrees of freedom are also included
in the interaction of the core and the target, so that dynamical
effects can occur during the reaction. We present results for the
breakup of $^{17}$C=$^{16}$C+n and $^{11}$Be=$^{10}$Be+n
on  $^{9}$Be.
Results show that the total cross section increases with
core deformation. More importantly, the relative percentage of the various
components of the initial state are modified during the reaction
process through dynamical effects.
This implies that comparing spectroscopic factors from structure calculations
with experimental cross sections requires more detailed reaction models that go
beyond the single particle model.
\end{abstract}


\maketitle

\section{Introduction}
\label{intro}

Over the last decade, technological developments have provided the
Radioactive Beam community with detailed data on a large variety of reactions
involving nuclei far from stability \cite{review-exp}.
Amongst these are breakup reactions, from which properties of loosely bound
systems can be extracted \cite{review-th}.
Breakup reactions have been used to determine ground state properties,
including binding energies, angular momentum structure,
momentum distributions and spectroscopic factors
(e.g.\ \cite{c19nakamura,aumann00,madda01,madda02,kelley95,li11orr,be11fukuda}),
as well as for studying excited state properties and resonant states (e.g.\ \cite{li10young}).

Breakup is not only of interest for studying the structure of
these nuclei, but also for astrophysical reasons. In a Coulomb dominated process,
if we can assume a one step semi-classical approximation \cite{alder75},
it is possible to factorize the Coulomb
dissociation cross section of $P+t \rightarrow (c+v)+t$ into a kinematical factor
multiplying the photo-dissociation cross section of the projectile
$c+v\rightarrow P$,  which in turn is easily translated into
the capture S-factor at  arbitrarily low energies between the fragment and core \cite{baur03}.
Many applications of this method have been used in the past
\cite{o14moto,kikuchi97,davids98} and new applications are
being developed \cite{fleurot05}.

In general, we need to improve on the semiclassical models, but still some approximations
need to be made,
depending on the energy regime, the target, and the
specific kinematical conditions of the experimental setup.
Recent results show that nuclear contributions extend
far out into regions where naively one expects they would be
negligible, because of the extended nature of loosely bound wave functions.
Coulomb-nuclear interference is also not negligible,
the far-field approximation of the Coulomb field is inaccurate,
and polarization effects need to be considered (see for example
\cite{esbensen96,dasso98,nunes98,esbensen99,nunes99,capel05}).
Even when there are no other bound excited
states, the breakup process goes through several continuum states
under the influence of the target. These effects have been referred to
as dynamical effects or continuum-continuum couplings, and together
with other improvements, have helped to solve apparent
discrepancies between different experiments \cite{ogata05,esbensen05}.

In earlier days, measurements involving
dripline nuclei detected only one of the fragments (e.g.\ \cite{kolata01}).
Those inclusive data contained inelastic breakup and transfer (stripping) as well as
elastic breakup (diffraction). In the latter, the projectile breaks up
and both fragments survive after the process without any excitation of
the target. In stripping, one of the fragments in the projectile gets
absorbed and/or the target gets excited. The improved beam
intensities in most facilities and the new phase of electronics in data acquisition
has allowed efficient coupling of several types of detectors, such that,
nowadays, data are less integrated and often contain complete kinematics
(e.g.\ \cite{marques00,gsi}).
For this reason, theoretical models
need to move toward disentangling the various process in their predicted
observables. In this work we will focus on the elastic breakup component.

Generally, breakup reactions have been described within a three body model,
where the projectile is simplified to a two body system ($P=c+v$).
The relative motion between the core and fragment of the projectile
is distinguished from, and then coupled to,
the relative motion of the projectile and the target.
In the high energy regime, straight-line trajectory or adiabatic
approximations are often made \cite{supa98,garrido99}.
At lower energies, semi-classical methods taking a
Rutherford orbit for the projectile's trajectory
are a possible choice \cite{dasso98,esbensen96,capel05}.
Over all energy regimes, one of the most successful models has been
the Continuum Discretized Coupled Channels method \cite{cdcc-theory}. It includes
the scattering states of the projectile completely coupled and is fully
quantal. In this model, nuclear and Coulomb are treated on equal footing,
thus interference effects appear naturally.
Also, any relevant multi-step process within the continuum,
or from the ground state to the continuum,
are automatically incorporated \cite{nunes99,tostevin01prc}.

As successful as it has been, CDCC \cite{cdcc-theory} has several limitations.
One of the most serious is the restriction to projectiles that can be approximated by two bodies
in relative motion.
The Kyushu group \cite{matsumoto04} has been expanding the original CDCC method
to solve problems where the projectile is of a clear three body
nature, such as Borromean systems ($^{6}$He, $^{11}$Li, $^{14}$Be, etc.).
Their method introduces a Gaussian expansion
with a complex argument \cite{egami04}, essential for computational efficiency.
Results for nuclear and Coulomb breakup have been recently
presented \cite{kamimura05,matsumoto05}. Another limitation is associated
with the non relativistic description, and this approximation becomes inaccurate at
very high energies.
Exploratory work solving the Klein-Gordon equation \cite{bertulani05} has
shown there to be non-negligible relativistic effects in the Coulomb dissociation
of $^8$B for experiments at GSI energies.
Finally, the projectile-target two-body asymptotics, as an approximation of the
pure three body asymptotics, has also been pointed out as a limiting
factor in certain reactions \cite{alt05}.

The CDCC method has been applied to many recent examples
\cite{rusek00,tostevin01prc,moro03,takashina03,summers04,ogata05}.
For the $^8$B breakup measured at Notre Dame where only the $^7$Be fragment was
detected \cite{kolata01}, the theoretical predictions agree very well with the data
for the kinematic region where only elastic breakup is expected \cite{tostevin01prc}.
If the same $^8$B structure model is used for the breakup of $^8$B at higher energies
\cite{davids01}, one needs to artificially adjust the quadrupole
strength of the couplings potentials in the reaction \cite{mortimer02}. This adjustment
cannot be understood by a simple renormalization of the $E2$ strength
in the $^7$Be-p continuum \cite{summers05} and suggests that structure beyond
the single particle $p_{3/2}$ is required to
describe the reaction  mechanisms at higher energies.
The elastic and inelastic scattering of protons on $^{11}$Be
in inverse kinematics was examined in the inert core single particle model \cite{shrivastava04}.
They found that serious discrepancies between the theory and data existed
in the $^{11}$Be\{$^{10}$Be(0$^+$)+n\} breakup spectrum in the region around
resonances built on a 2$^+$ core.

There are several ways to improve the  two body single particle
description of the projectile. While the Kyushu group has focused on
the description of breakup of a three body projectile, in this work
we will discuss a generalization of the CDCC method for a two body projectile
where the wave function is no longer single particle.
This generalization of the CDCC method, which we denominate
eXtended CDCC (XCDCC) \cite{summers06}, includes a coupled channels
description of the projectile's initial and final states, but also any dynamical core
excitation/de-excitation that may occur in the reaction process.

It is clear that if core excitation/de-excitation happens during the
reaction process,
this will have implications for not only the excited cross sections,
but also for measuring the ground state core distributions.
In addition, one expects that the excitation/de-excitation process will depend
strongly on the type of reaction, namely high versus low energy,
light versus heavy target.
Even when there is only a single valence neutron/proton, the
valence nucleon may occupy several configurations, in some of which the core
is excited. By measuring the gamma rays in coincidence
with outgoing fragments, core states can be identified, and
spectroscopic factors for the overlap of the projectile ground state with specific core states
can be examined \cite{knockout}.

In nuclear dominated knockout reactions on light targets,
typically the method of \cite{tostevin99} is followed.
Therein one calculates the single particle cross section for both diffraction (elastic breakup)
and stripping, in an eikonal model, for each ground state configuration of the projectile.
The single particle cross sections for each state are then multiplied by the spectroscopic factor
for that configuration, calculated within a shell model, and the total knockout cross sections
is obtained by adding the single particle contributions incoherently.
Improvements on the eikonal approximation for the breakup contribution
\cite{tostevin01npa,tostevin02} have been examined using CDCC.
These models, however, are still single particle and neglect any
interference between configurations, as well as any dynamical
excitation of the core in the reaction.
We will examine the importance of these effects in this paper.

In Coulomb dissociation, by contrast, a more simplistic approach is typically assumed:
that the relative probabilities do not change during the reaction,
and the contribution from excited core components is negligible \cite{be11fukuda}.
Thus the ratio of the theoretical cross section, for the ground-state-core configuration,
to the experimental cross section is often taken to be
directly the spectroscopic factor for the ground state.

Contributions from excited core components in Coulomb dissociation have previously been estimated
\cite{shyam01}, by calculating the contribution from excited core components separately,
and adding them incoherently, following the method of \cite{tostevin99}.
The Coulomb dissociation was calculated using DWBA and Adiabatic reaction models.
In \cite{shyam01}, it was concluded that the Coulomb dissociation to excited core states
was very small, only a few percent, very much less than in nuclear breakup.
They then suggest that since the cross section from excited core states has a negligible contribution to the total Coulomb breakup cross section,
one could compare the single particle theoretical cross section to the experimental cross section
to obtain the spectroscopic factor for the core in its ground state.
One has to consider the many approximations assumed in
the theory.
This model neglected any interference between the configurations and any
dynamical (de-)excitation of the core.
These effects can be examined with XCDCC, but Coulomb dominated reactions
will be a subject of a future publication.

One simple way to introduce collective degrees of freedom to the core is
to assume it behaves as a perfect rotor.
\emph{Rotor}+\emph{nucleon} coupled channels models for loosely
bound systems have been applied to a number of light nuclei on the
dripline (e.g.\ \cite{be11nunes,be12nunes,ridikas98}).
This type of structure model for
the projectile has been also used in the calculations of reaction
observables  \cite{tarutina04,batham05}.

In \cite{tarutina04}, the coupled wave functions are
used to compute the B(E1) electric excitation function from the
ground state into the continuum. The interference between the
states with the same core spin in a pure Coulomb dissociation process is then studied within
the semi-classical first order approximation \cite{tarutina04}.
Interference effects between states with different core spins were neglected
since this model does not include inelastic transitions caused by the target.
These interference effects were shown to be important
in $^{17}$C on $^{208}$Pb at 67 A MeV.

In \cite{batham05}, the eikonal approximation
is used to calculate the total cross sections for the
breakup of a deformed core plus nucleon system. Results for $^{17}$C breakup
on $^9$Be at 62 A MeV show a dramatic  increase of the cross section with the
quadrupole deformation parameter. Contrary to the semi-classical first order
approximation, this model includes dynamical effects,
but only total cross sections are obtained. It can be interpreted as a
precursor to the new model presented here.

There is a wide variety of problems that can be investigated with XCDCC,
and this is the first chapter where the formalism is given in detail and integrated observables for two cases are presented, namely $^{17}$C  and $^{11}$Be breakup
on $^{9}$Be at $\approx 60$ A MeV, both measured at MSU \cite{aumann00,madda01}.
We present a comparison with the data of \cite{aumann00} in a shorter paper \cite{summers06}.

In section \ref{theory} the theoretical development is presented.
A description of the structure models used for the projectiles is detailed
in section \ref{structure}.
Numerical details for the examples considered are given in section \ref{numerical}.
Following, in section \ref{results}, the results are detailed and discussed.
Finally, section \ref{conclusions} contains the conclusions drawn from our study and
an outlook for future work.

\section{\label{theory} Theoretical developments}

\begin{figure}
\includegraphics[width=6cm]{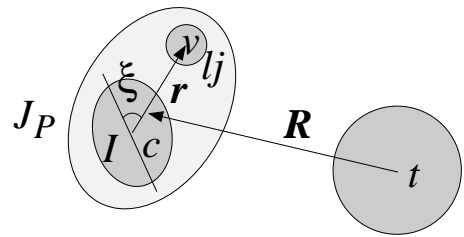}
\caption{\label{FIG:coord} Coordinates for the three body scattering problem to be
considered.}
\end{figure}

The breakup of a projectile $P$ consisting of a valence particle ($v$) loosely bound to a core
($c$) on a target $t$ can be modelled as a three body
scattering problem. The three body Hamiltonian for the system has the form:
\be\label{H-cdcc}
H = \TR + H_{\mathrm{proj}} + V_{ct} + V_{vt} \ ,
\ee
where $\TR$ represents the kinetic energy operator for the projectile-target system,
$H_{\mathrm{proj}}$ is the internal Hamiltonian of the projectile
and $U_{ct}$ ($U_{vt}$) is the interaction between the core (valence) particle and the
target.
The Hamiltonian is usually written in the Jacobi coordinates $(\vec{R},\vec{r})$ where
$\vec{R}$ is the coordinate for the center of mass of the projectile relative to the
center of mass of the target, and $\vec{r}$ is the coordinate of the valence
particle relative to the center of mass of the core as shown in
\fig{FIG:coord}.

The full three body wave function for the system with definite total spin $J_T$
is expanded over the states of the projectile, both bound and unbound:
\be\label{wf3b}\begin{split}
&\Psi_{J_T}^{M_T}(\coords)  \\ &=\sum_{\alpha} \Psi^{J_T}_{\alpha}(R)
\big[[ Y_L(\hat{\vec{R}}) \otimes \Phi_{J_P}^{in}(\vec{r},\xi) ]_J \otimes
   \Phi_{J_t}(\xi_t) \big]_{J_T}^{M_T} \ .
\end{split}\ee
The XCDCC states, labelled by $\alpha\equiv\{L,J_P,J,J_t,i,n\}$,
describe the projectile state $\{J_P,i,n\}$,
where $J_P$ is the total spin of the projectile,
$i$ is the discretization in relative momentum and $n$ is an
additional boundary condition that is
required to describe coupled channels bins and is discussed later.
In this paper we assume the target is structureless, therefore removing the dependence on $(\xi_t)$ in \eq{wf3b}.
The projectile-target orbital angular momentum,
$L$, and the spin of the target $J_t$ all couple to a total spin of the three-body system
$J_T$.
We use the notation
\be\label{coup-alpha}
\braket{\hat{\vec{R}},\vec{r},\xi}{\alpha;J_T} = \big[[ Y_L(\hat{\vec{R}})
\otimes \Phi_{J_P}^{in}(\vec{r},\xi) ]_J \otimes \Phi_{J_t} \big]_{J_T}
\ee
to describe the angular momentum and projectile wave function compactly.

If the interaction between the fragments in the projectile depends only on $\vec{r}$,
the projectile Hamiltonian $H_{\mathrm{proj}}$ generates single particle bound state
wave functions,
and single particle continuum states which can later be discretized into continuum bins.
In this work, we wish to include an additional degree of freedom, $\xi$, concerning
the projectile's core, and allow the core states to be coupled together.
For simplicity we take a rotational model for the core \cite{be11nunes},
but the evaluation of the coupled channels bins and potential matrix elements
are not specific to the choice of this collective model.
We consider that the core-valence interaction depends on core collective
degrees of freedom $\xi$:
\be\label{hproj}
H_{\mathrm{proj}} = \Tr + V_{vc}(\vec{r}, \xi) + h_{\mathrm{core}}(\xi) \; ,
\ee
where $\Tr$ is the core-valence kinetic energy operator, and
$h_{\mathrm{core}}(\xi)$ is the intrinsic Hamiltonian for the core. We define
$\varphi_{I}$ to be the eigen function of the core corresponding to
eigen-energies $\varepsilon_{I}$. Now the projectile ground state contains
contributions from different core states, in a fully coupled manner:
\be\label{wfproj}
\Phi_{0}(\vec{r},\xi) = \sum_{a} \phi_{a}(r)
\big[[ Y_l(\hat{\vec{r}}) \otimes \chi_s ]_j  \otimes  \varphi_{I}(\xi) \big]_{J_P}
\ee
We use the subscript 0 on the projectile wave function to denote the ground state (the supercripts $(i,n)$ in \eq{coup-alpha} are superfluous for bound states).
Here, a given channel $a$ depends on a set of quantum numbers describing the
relative ($l$) and total ($j$) angular momentum of the valence particle,
with fixed spin $s$, which couples to the core spin $I$.
We label each channel which can couple to total spin $J_P$ by $a\equiv\{l,j,I\}$.
We use the notation
\be\label{coup-jp}
\braket{\hat{\vec{r}},\xi}{(ls)j,I;J_P} = \big[[ Y_l(\hat{\vec{r}}) \otimes \chi_s ]_j
\otimes  \varphi_{I}(\xi) \big]_{J_P}
\ee
to describe the angular momentum and core wave function compactly.
Previous work \cite{be11nunes} gives details of such \emph{rotor}+n models.

In addition to the dependence of the internal Hamiltonian of the projectile on
$\xi$, the core-target interaction $V_{ct}(\vec{R},\xi)$ also contains the
dependence on the degrees of freedom of the core and therefore, for a breakup
process within this model, the core-target interaction can excite/de-excite the
core during the reaction process. The coupling matrix elements of this deformed
potential will be discussed in section~\ref{couplings3b}.

\subsection{\label{coupledbins} Construction of projectile coupled channels bins}

The projectile's ($c$+$v$) scattering states for a given incoming
kinetic energy $E_k$ are solutions of a coupled channels equation which can be
reduced to the following radial form:
\be\label{hccbin}
[E_k - \varepsilon_{a}
+ \varepsilon_{n} - T_a] f_{a:n}(r;k_{an}) = \sum_{a'} V_{a:a'}(r)
f_{a':n}(r;k_{a'n}) \ ,
\ee
where $a$ and $n$ stand for outgoing and incoming boundary conditions respectively,
$E_k = \hbar^2 k^2/2\mu_{vc}$ is the core-valence relative energy, $k$ is the core-valence
relative momentum, and the reduced mass is $\mu_{vc}$.
The momentum that appears in the scattering wave, $k_{an}$,
is the asymptotic momentum available in particular channel $an$, and is defined by
\be
k_{an}^2 = k^2 - \conkvc [\varepsilon_a-\varepsilon_n] \ ,
\ee
where $\varepsilon_{a}$ is the energy of the core in the outgoing channel $a$,
and $\varepsilon_{n}$ is the energy of the core in the incoming channel, $n$.
The kinetic energy operator,
\be
T_a = -\convc \left[ \dderiv{r} - \frac{l(l+1)}{r^2} \right] \ ,
\ee
is defined for the angular momentum $l$ of the core-valence system in channel $a$.
The coupling potential matrix is
\be\label{bincoup}
V_{a:a'}(r) = \bra{(ls)j,I;J_P} V_{vc}(\vec{r},\xi)
\ket{(l's)j',I';J_P} \ ,
\ee
where $a\equiv\{l,j,I\}$ are the outgoing channel quantum numbers.

The open scattering wave functions are obtained by matching to the asymptotic boundary
condition,
\be\label{fasymp}
f_{a:n}(r;k) = \frac{\im}{2}[H^{-}_n(kr)\delta_{an} - H^{+}_a(kr) S_{a:n}^{J_P}]\ ,
\ee
at $r_\mathrm{match}$ to obtain the coupled channels square $S$-matrix, $S_{a:n}^{J_P}$.
Here $H^{(\pm)}_a(kr)$ are the diagonal Coulomb functions.
Inelastic channels with excited core components will remain closed unless the relative energy
is higher than the core energy or the incoming channel has an excited core component.
If the available energy for channel ($an$) is negative, then the wave function $f_{a:n}(r;k_{an})$
will be closed with separation energy ${\hbar^2k_{an}^2}/{2\mu_{vc}}$.

In principle, coupled channels scattering wave functions could be used directly to calculate
the one-step breakup process. However, as multi-step effects can be extremely important,
we opt to discretize the coupled channels set, into square integrable bin wave functions.
The generalization of the standard binning procedure \cite{cdcc-theory}
to coupled channels bins is described below.

For each projectile $J_P$ state in the continuum,
we define coupled channels bin wave functions that are square integrable, resulting from
a superposition of pure scattering states (calculated from \eq{hccbin}) within a
momentum bin $[k_{i-1},k_{i}]$. Each momentum interval is labelled by $i$.
The coupled channels bin wave function are written as
\be\label{ccbin}
u^i_{a:n}(r) = \sqrt{\frac{2}{\pi N^i_{n}}} \int_{k_{i-1}}^{k_i} \dif k ~ g_{n}(k)
f_{a:n}(r;k_{an})
\ee
where, as before, $k$ is the relative momentum between core and valence fragments.
Here, $k_0$ represents the breakup threshold between the core and valence particles.

The weight factor of the coupled channels bins, $g_{n}(k)$, is chosen to be
$\exp{-\im\delta_{n}(k)}$,
where $\delta_{n}(k)$ is the phase shift of channel $a=n$
(i.e.\ the phase obtained from the diagonal of the $S$-matrix).
In the one channel case, this phase makes the wave functions real \cite{nunes99}.
The coupled channel bins are now complex, and the full complex wave function
is used in the CDCC equations up to the matching radius $r_\mathrm{match}$.
By choosing the weight factor as the phase of channel $a=n$, the wave functions are approximately real.
This is useful when matching to the asymptotic Coulomb functions \cite{crcwfn} in the CDCC equations,
as the long-range Coulomb functions become approximately real.

The normalization factor is
\be
N^i_{n} = \int_{k_i-1}^{k_i} | g_{n}(k) |^2 \dif k  \ .
\ee
Since our weighting has $|g_{n}(k)|^2=1$, the
normalization is simply $N^i_{n}=k_i-k_{i-1}$, and the average energy of bin $\alpha$ is
$\hat\varepsilon_{\alpha} = \hbar^2 (k_i^3 - k_{i-1}^3) / 6 \mu_{vc} (k_i-k_{i-1})$
\cite{tostevin01prc}.

The full bin wave function describing the projectile state $J_P$, for momentum
bin $i$ and incoming wave $n$,
is the sum of the coupled channels bins obtained from \eq{ccbin}:
\be\label{projwf}
\Phi_{J_P}^{in}(\vec{r},\xi) = \sum_{a} u^i_{a:n}(r) \braket{\hat{\vec{r}},\xi}{(ls)j,I;J_P} \; .
\ee
The incoming channels form an orthogonal set and therefore projectile wave functions carry the
label $n$.
It is important to note that since the bin wave functions appear in the bra in
the three-body $T$-matrix, these incoming wave boundary conditions on the scattering wave
relate to the outgoing channel of the breakup fragments.

\subsection{\label{couplings3b} The three body coupling potentials}

The scattering problem for the three body Hamiltonian, \eq{H-cdcc} $H \Psi =E \Psi$,
can be reduced to a set of radial coupled channels equations in $R$ when using
the expansion \eq{wf3b}. Then the scattering coupled channels equation becomes
\be
[E_{\alpha} - T_L - U^{J_T}_{\alpha:\alpha}(R)] \Psi^{J_T}_{\alpha}(R) =
\sum_{\alpha'\ne\alpha} U^{J_T}_{\alpha:\alpha'}(R) \Psi^{J_T}_{\alpha'}(R) \; ,
\ee
where the kinetic energy operator is
\be
T_L = -\conpt \left[ \dderiv{R} -\frac{L(L+1)}{R^2} \right] \ ,
\ee
and $\mu_{Pt}$ is the reduced mass of the projectile-target system.
The energy of each channel $E_{\alpha}=E-\hat\varepsilon_{\alpha}$ is the total energy of
the three-body system minus the average energy of bin $\alpha$.
The incident center-of-mass energy will be $E_{0}$, as $\alpha=0$ represents the ground state
of the projectile.

The coupling potentials for the elastic breakup of the projectile (the process where the
target remains in its ground state) are
\be\begin{split}\label{coupling-pot}
U^{J_T}_{\alpha:\alpha'}(R) = \bra{\alpha;J_T} V_{ct}(\coords) + V_{vt}(\vec{R},\vec{r})
\ket{\alpha';J_T} \;,
\end{split}\ee
where $\alpha$ describes the angular momentum as defined in \eq{coup-alpha}.
The projectile-target interaction potential is the sum of the core-target ($V_{ct}$)
and valence-target ($V_{vt}$) potentials, and contains both nuclear and Coulomb terms.
The valence-target interaction can be treated as before \cite{fresco}, by the theory of Appendix
\ref{qzero}. The new physical element is in the core-target interaction, which now depends on the
core degrees of freedom ($\xi$).

The core-target interaction can be expanded in multipoles of $Q$,
\be\label{expansionQ}
V_{ct}(\coords) = \sum_Q \hat{Q}^2 V^Q_{ct}(\vec{r_c}) \sum_q C_{Qq}(\vec{\hat{r_c}})C^*_{Qq}(\xi)
\ee
where $\vec{r_c}=\vec{R}-\gamma\vec{r}$ is the core-target coordinate,
with $\gamma=m_v/m_P$ as the ratio of the valence mass to the projectile mass,
and $C_{Qq}(\vec{\hat{r_c}})$ is  related to the spherical harmonic,
\be
Y_{Qq}(\vec{\hat{r_c}}) = \hat{Q} C_{Qq}(\vec{\hat{r_c}}) /\sqrt{4\pi} \ ,
\ee
where we use the notation $\hat{Q}=\sqrt{2Q+1}$.

Next, the product $r_c^Q C_{Qq}(\vec{\hat{r_c}})$ can be expanded in multipoles
$\lambda$ for the $\vec{r}$ and $\vec{R}$ coordinates, as in \cite{brink}:
\be\begin{split}\label{expansionlambda}
r_c^Q C_{Qq}(\vec{\hat{r_c}}) =& \hat{Q}^3 \sum_{\lambda\mu}
\sqrt{\frac{(2Q)!}{(2\lambda)![2(Q-\lambda)]!}}
R^{\lambda} (-\gamma r)^{Q-\lambda} \\ &\times
C_{\lambda \mu}(\vec{\hat{R}}) C_{Q-\lambda, q-\mu}(\vec{\hat{r}})
\threej{Q-\lambda}{\lambda}{Q}{q-\mu}{\mu}{-q} \ .
\end{split}\ee

The multipole $Q$ for the core-target potential $V^Q_{ct}(\vec{r_c})$ can be
expanded in multipoles of $K$,
\be\label{expansionK}
\frac{V^Q_{ct}(\vec{r_c})}{r_c^{Q}} = \sum_K \hat{K}^2 V^{QK}_{ct}(r,R)
\sum_k C_{Kk}(\vec{\hat{R}})C^*_{Kk}(\vec{\hat{r}}) \ ,
\ee
where the radial potential dependence is given by
\be
V^{QK}_{ct}(r,R) = \frac{1}{2} \int_{-1}^{+1} \frac{V^Q_{ct}(\vec{r_c})}{r_c^{Q}} P_K(u)
\dif u\quad; \quad u=\vec{\hat{R}}\cdot\vec{\hat{r}} \ .
\ee
Combining Eqs.~(\ref{expansionQ}), (\ref{expansionlambda}) and (\ref{expansionK}),
we find that the spherical harmonics in $\vec{R}$ can be coupled to a new multipole $\Lambda$
and, similarly, the spherical harmonics in $\vec{r}$ can be coupled to another multipole $\Lambda'$.
The angular momentum structure gives three spherical harmonics in  $\vec{\hat R},\vec{\hat r},\xi$
coupling to form a tripolar spherical harmonic of rank zero.
Summing over the angular momentum projections $(q,\mu,k)$,
the core-target potential operator can then be reduced (according to Appendix
\ref{details}) to
\bwt
\be\begin{split}\label{vct-coupling}
V_{ct}(\coords) =& \sum_{KQ\lambda} V^{QK}_{ct}(r,R) R^{\lambda}
(-\gamma r)^{Q-\lambda} \sqrt{\frac{(2Q)!}{(2\lambda)![2(Q-\lambda)]!}}
(-1)^{Q} \hat{Q}\hat{K}^2 \\ &\times
\sum_{\Lambda\Lambda'} \hat{\Lambda}\hat{\Lambda'}^2
\threejz{K}{\lambda}{\Lambda} \threejz{K}{Q-\lambda}{\Lambda'}
\sixj{\Lambda'}{\Lambda}{Q}{\lambda}{Q-\lambda}{K}
\tripolar \;.
\end{split}\ee
\ewt

The transition couplings of \eq{coupling-pot} are calculated as
matrix elements of \eq{vct-coupling}, assuming the
coupling order defined in \eq{wf3b}.
It is useful to write the transition couplings in terms of a
\emph{transition potential} which only depends on the overall
projectile initial and final states,
$J_P',i',n'$ and $J_P,i,n$, and the transition multipolarity $\Lambda$.
Combining the potential operator from \eq{vct-coupling} and the matrix elements from
Appendix \ref{details}, we write the core-target three-body couplings as
\be\begin{split}\label{vjt}
U^{J_T}_{\alpha:\alpha'}(R) =&
\hat{L}\hat{L'}\hat{J_P}\hat{J_P'}
(-1)^{J_P+J} \sum_{\Lambda} (-1)^{\Lambda} \hat{\Lambda}^2 \\ &\times
\threejz{\Lambda}{L}{L'} \sixj{J_P}{J_P'}{\Lambda}{L'}{L}{J}
F^{\Lambda}_{J_Pin:J_P'i'n'}(R) \ .
\end{split}\ee
The form factor is the sum over $KQ\lambda$ multipoles and the projectile
coupled channels ($a$ and $a'$), \be\label{formf}
F^{\Lambda}_{J_Pin:J_P'i'n'}(R) = \sum_{KQ\lambda\atop aa'}
R^{KQ\lambda}_{ain:a'i'n'}(R) P^{KQ\lambda:\Lambda}_{a:a'} \ ,
\ee
where the radial dependence is defined by the integral:
\be\begin{split}\label{radial}
&R^{KQ\lambda}_{ain:a'i'n'}(R) \\ & =\hat{K} \int_0^{R_m}
u^{i}_{a:n}(r)^* V^{QK}_{ct}(r,R)  R^{\lambda}
(-\gamma r)^{Q-\lambda} u^{i'}_{a':n'}(r) \dif r \ .
\end{split}\ee
The remaining angular momentum dependence is summarized as
\bwt
\be\begin{split}\label{pcoup}
P^{KQ\lambda:\Lambda}_{a:a'} =& (-1)^{j'+l+l'+s+Q}
\hat{Q}\hat{K}\hat{j}\hat{j'}\hat{l}\hat{l'}
\sqrt{\frac{(2Q)!}{(2\lambda)![2(Q-\lambda)]!}}
\threejz{K}{\lambda}{\Lambda} \reduced{I}{C_{Q}(\xi)}{I'}
\\ &\times
\sum_{\Lambda'} \hat{\Lambda'}^2
\threejz{K}{Q-\lambda}{\Lambda'} \threejz{\Lambda'}{l}{l'}
\sixj{\Lambda'}{\Lambda}{Q}{\lambda}{Q-\lambda}{K}
\sixj{j}{j'}{\Lambda'}{l'}{l}{s}
\ninej{J_P}{J_P'}{\Lambda}{j}{j'}{\Lambda'}{I}{I'}{Q}\; .
\end{split}\ee
\ewt

The angular momentum coupling in \eq{pcoup} contains the matrix element
$\reduced{I}{C_{Q}(\xi)}{I'}$, which describes elastic and inelastic excitations
of the core through its scattering by the target.
In this paper we use a rotational model for the core, for which this matrix element
has a simple solution (Appendix \ref{rotor}).

This matrix element $\reduced{I}{C_{Q}(\xi)}{I'}$ is left separate in \eq{pcoup} since,
in general, other models could be used to induce the inelastic scattering of the core.
Given a core-target interaction with core degrees of freedom, $\xi$, and
multipole expansion, $Q$, the inelastic modes can be calculated, and along with
radial coupling potentials [\eq{bincoup}] for the core-valence interaction,
XCDCC opens the door for other models of the core which go beyond the collective
model.

The monopole couplings for the core-target interaction
[$Q=0$ in Eqs.~(\ref{vjt}-\ref{pcoup})] are evaluated in Appendix~\ref{qzero}.
We see by comparing \eq{vjt} and \eq{coup-q0} that including extra core degrees of freedom
for the core does not increase the complexity of the final form factors, since they
both depend only on a multipolarity and the transition $J_P' \rightarrow J_P$.
Thus the form factors can be pre-calculated efficiently. To reduce memory
requirements, the individual components $R^{KQ\lambda}_{ain:a'i'n'}(R)
P^{KQ\lambda:\Lambda}_{a:a'}$ can be summed so only
the final $F^{\Lambda}_{J_Pin:J_P'i'n'}(R)$ stored.
A new version of {\sc Fresco} \cite{fresco} was created for this purpose.

The only increase in the size of the calculation compared with the single particle calculations,
with no core degrees of freedom, is the increase in the model space due to the extra
quantum number $n$ for the projectile state.
This relates to the incoming wave boundary condition on the
coupled channels solution of the core-valence wave function.
One needs to include all possible incoming waves for each of the coupled channels scattering states.
This extra quantum number carries the information regarding the final state of the
core-valence system in the CDCC equations.

\section{\label{structure} Structure of $^{11}$Be and $^{17}$C in the rotational model}

We use a rotational model to describe the projectile as a coupled core-valence system.
The rotational model for $^{17}$C is based on the model used in \cite{ridikas98}.
We use the same potential parameters as \cite{ridikas98},
given in Table~\ref{TABLE:c17-pot}.
We neglect the spin of the valence particle,
because we are interested in the effect of the core spin, and the
effect of the valence particle spin on the total cross sections is small.
This model produces a $^{17}$C ground state spin of $J_P^\pi=2^+$.
We use a quadrupole mass deformation of $\beta_2=0.55$, as in \cite{batham05}.
Modeling the $^{16}$C core with two states $\{0^+,2^+\}$, with an excitation energy of
$E_{2^+}=1.766$ MeV, and including all possible relative angular momentum up to
$l=4$, produces four coupled channels in the ground state: $[d\otimes0^+]$,
$[s\otimes2^+]$, $[d\otimes2^+]$, $[g\otimes2^+]$, with relative probabilities
of 0.175, 0.103, 0.721, 0.001 respectively. This model predicts the
dominant channel to be $[d\otimes2^+]$, which is consistent with the large
spectroscopic factor obtained from shell model calculations (this channel was
left out of the model space in the rotational model of \cite{ridikas98}). The
single particle (SP) central potential parameters are taken from
\cite{batham05}, and are adjusted to fit the ground state binding energy for
each configuration. $^{16}$C also has other bound states around 4 MeV in
energy, one of which, the $4^+$ at $E_{4^+}=4.142$ MeV, could be included within
our rotational model.
In some of the calculations, we include this core
state in the projectile continuum although we keep the projectile's ground
state description without the $4^+$ core state for consistency. Using a rms
matter radius of 2.70 fm \cite{ozawa01} for the $^{16}$C core, the rms radius
calculated for the deformed $^{17}$C is 2.77 fm. The rms radius for the single
particle $[d\otimes2^+]$ state is 2.78 fm. The weighted sum for the
incoherently summed single particle projectile states, has a rms radius of 2.79 fm.
The experimental determined rms radus is 3.04$\pm$0.11 fm \cite{stlaurent89}.

\begin{table}\begin{tabular}{|c|c|cccc|}\hline
model & interaction                 & $V$        & $R_V$     & $a_V$     & $\beta_2$ \\ \hline
CC    & $^{16}$C\{$0^+,2^+$\}+n     &  62.0      & 3.2       & 0.65      & 0.55      \\
SP    & $^{16}$C\{$2^+$\}+n\{$d$\}  &  52.7      & 3.2       & 0.7       & -         \\
SP    & $^{16}$C\{$2^+$\}+n\{$s$\}  &  49.4      & 3.2       & 0.7       & -         \\
SP    & $^{16}$C\{$0^+$\}+n\{$d$\}  &  48.5      & 3.2       & 0.7       & -         \\ \hline
\end{tabular}
\caption{\label{TABLE:c17-pot}$^{16}$C+n potential parameters for \eq{ws-pot}.}
\end{table}

For $^{11}$Be we base our potential parameters on those of
\cite{be11nunes}.
These were fitted to give the correct binding energies of the
ground and first excited state, as well as the B(E1) transition between them,
using a parity dependent potential.
Since again we neglect the spin of the neutron, and therefore no spin-orbit force,
we readjust the potential parameters slightly to fit the binding energies of the bound
states.
The parameters used in the calculation are given in
Table~\ref{TABLE:be11-pot}, with the model CC referring to the coupled
rotational model for the projectile, where a deformation $\beta_2=0.67$
is taken from \cite{be11nunes}.
This model produces a $^{11}$Be ground state
with spin $J_P=0^+$, with two coupled channels: a s-wave neutron coupled to
$0^+$ core, and a d-wave neutron coupled to a $2^+$ core.
The relative probabilities
are 0.883 and 0.117 for the $[s\otimes0^+]$ and $[d\otimes2^+]$ states
respectively. The single particle model for the core consists of a parity
dependent potential as used in \cite{be11nunes}, which fits the bound states
using a central potential. For the states with a $2^+$ core the potential depth
is adjusted to fit the binding energy plus the energy of the $2^+$ core. Using
a rms matter radius of 2.28 fm \cite{alkhalili96} for the $^{10}$Be core, the
rms radius calculated for the deformed $^{10}$Be+$n$ is 2.99 fm. The rms radius
for the single particle $0^+$ core calculation is 3.08 fm. The weighted sum
for the incoherently summed single particle projectile states has a radius
of 3.01 fm. These compare reasonably well to the rms radius for $^{11}$Be of 2.90$\pm$0.05
fm \cite{alkhalili96}.
The radii in our few-body model are a bit small, but since we do not compare to
data, and only compare to previous work, we keep the same potential model for consistency.

\begin{table}\begin{tabular}{|c|c|c|cccc|}\hline
model & interaction                 & parity    & $V$       & $R_V$     & $a_V$     & $\beta_2$ \\ \hline
CC    & $^{10}$Be\{$0^+,2^+$\}+n    & +         & 55.25     & 2.483     & 0.65      & 0.67      \\
CC    & $^{10}$Be\{$0^+,2^+$\}+n    & --        & 47.00     & 2.483     & 0.65      & 0.67      \\
SP    & $^{10}$Be\{$0^+$\}+n        & +         & 55.50     & 2.736     & 0.67      & -         \\
SP    & $^{10}$Be\{$0^+$\}+n        & --        & 30.48     & 2.736     & 0.67      & -         \\
SP    & $^{10}$Be\{$2^+$\}+n        & +         & 75.07     & 2.736     & 0.67      & -         \\
SP    & $^{10}$Be\{$2^+$\}+n        & --        & 39.95     & 2.736     & 0.67      & -         \\ \hline
\end{tabular}
\caption{\label{TABLE:be11-pot} $^{10}$Be+n potential parameters for \eq{ws-pot}.}
\end{table}

\section{\label{numerical} Numerical considerations}

In all the calculations the core-valence integrals were performed out to a maximum radius of
$r_\mathrm{match}=70$ fm, with a step length of $h=0.05$ fm.
The momentum integrations for each bin [\eq{ccbin}] were calculated with 60 steps.
The multipole transitions were calculated with all possible $K$, $Q$ and $\lambda$ transitions,
with the transition multipolarity truncated to $\Lambda\le 4$.
Partial waves were included up to $L=5000$ for $^{17}$C breakup and up to $L=3000$ for $^{11}$Be
breakup.
The radial wave functions were matched to the asymptotic Coulomb functions at a radius of
$R_\mathrm{asym}=500$ fm.
The spin of the valence particle was neglected and therefore the channel labels all have $j=l$.
The spin of the $^{9}$Be target was set to zero as it is a spectator in our reaction model.
The channel labels refer to the total spin and parity of the projectile ($J_P^\pi$), the
angular momentum of the neutron ($l$), and the spin of the core fragment ($I$), in the final state.

\begin{figure*}
\includegraphics[width=12cm]{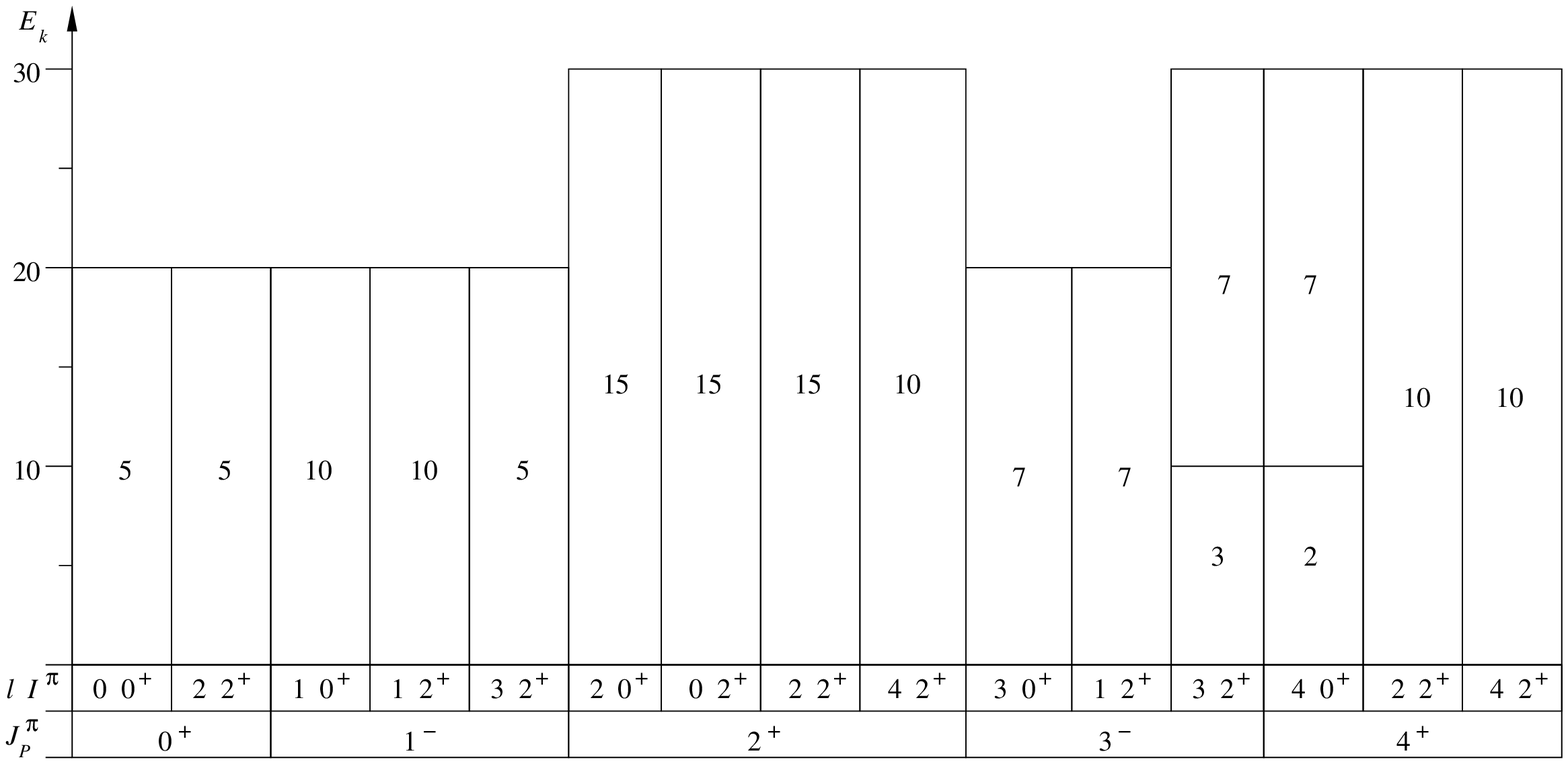}
\caption{\label{FIG:c17-2+-modelspace}$^{17}$C continuum model space.
The number of bins and the energy range are given for each outgoing channel ($l,I^\pi$) for each spin parity combination
of the projectile ($J_P^\pi$).}
\end{figure*}

\begin{figure*}
\includegraphics[width=18cm]{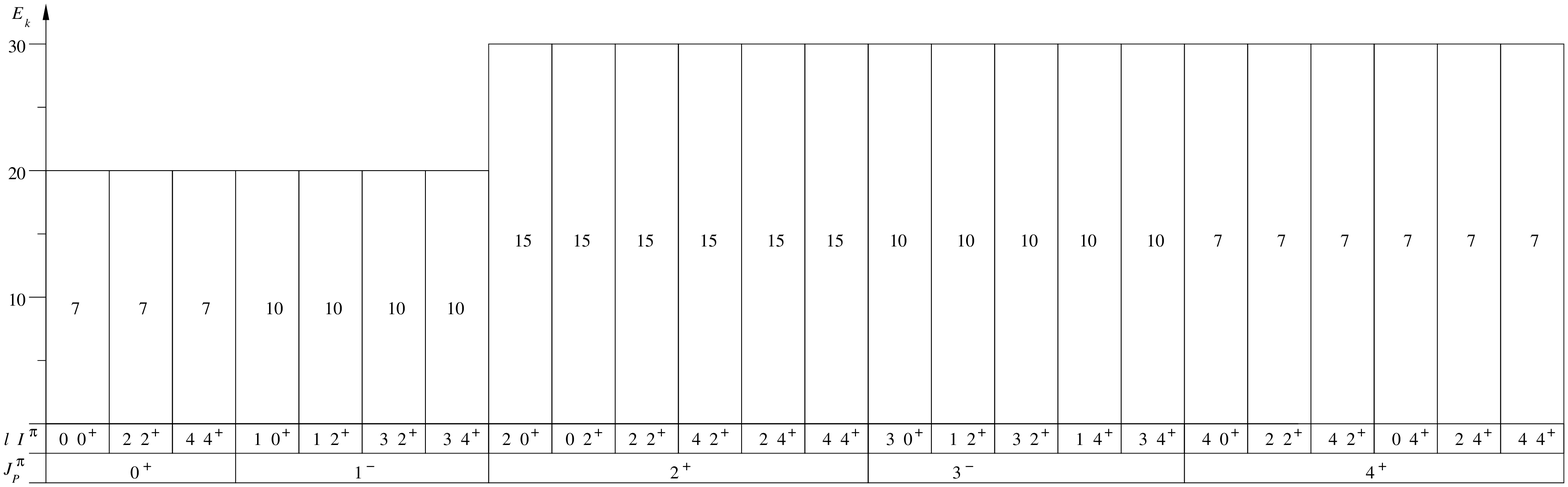}
\caption{\label{FIG:c17-4+-modelspace}$^{17}$C continuum model space.
The number of bins and the energy range are given for each outgoing channel ($l,I^\pi$) for each spin parity combination
of the projectile ($J_P^\pi$).}
\end{figure*}

\begin{figure*}
\includegraphics[width=12cm]{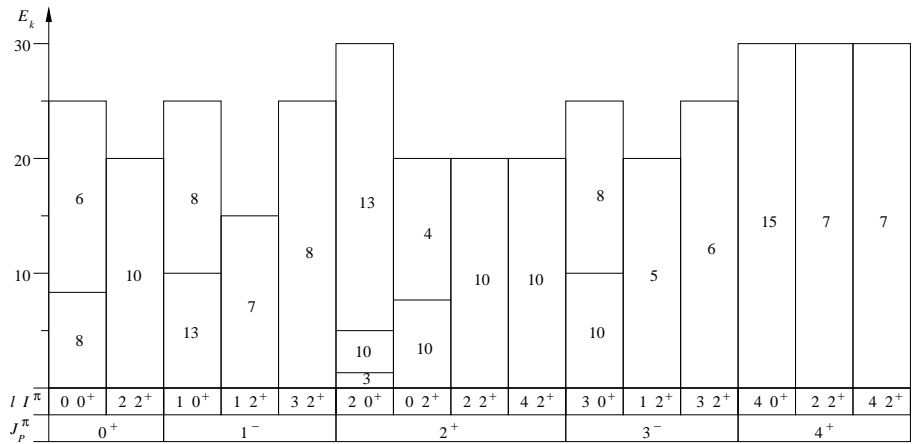}
\caption{\label{FIG:be11-modelspace}$^{11}$Be continuum model space.
The number of bins and the energy range are given for each outgoing channel ($l,I^\pi$) for each spin parity combination
of the projectile ($J_P^\pi$).}
\end{figure*}

The XCDCC model space for the $^{17}$C reaction where $^{16}$C is modeled
with two states $\{0^+,2^+\}$, the XCDCC model space is given in
\fig{FIG:c17-2+-modelspace}.
For the $^{17}$C reaction where $^{16}$C is
modeled with three states $\{0^+,2^+,4^+\}$, the XCDCC model space is given in
\fig{FIG:c17-4+-modelspace}.
The XCDCC model space for $^{11}$Be breakup is given in
\fig{FIG:be11-modelspace}.
The numbers of bins spanning an energy range for
each channel in the final state are given in each box, with
the bins at regular intervals in momentum. Summaries of
the various sizes for the model space are given in table~\ref{TABLE:size}. The
first column is the projectile model space: which core states are included
in the calculation. The next five columns give the number of permutations for
each of the variables; the number of coupled channel states ($aJ_Pin$), the
number of projectile states ($J_Pin$), the number of couplings between coupled
channel states ($\formp$), the number of form factors stored ($\formf$), and the
number of channels in the CDCC equations ($\alpha$). The deformed core
calculations were performed on a SGI Altix 3700, while the single particle
calculations were performed on a cluster of AMD Opteron 250 processors.

\begin{table*}\begin{tabular}{|c||ccccc|cccc|}\hline
Projectile                 & $aJ_Pin$  & $J_Pin$   & $\formp$  & $\formf$  & $\alpha$  & CPUtime(s) &  NCPUS  & walltime(h:m) & Mem(GB) \\ \hline
$^{10}$Be\{$0^+$\}+n       &        -  &        90 &        -  &     7312  &     258   &   20568    &     11  &     00:35     & 0.2     \\
$^{10}$Be\{$0^+,2^+$\}+n   &      572  &       179 &  3862133  &    58951  &     542   &  342008    &      8  &     13:32     & 1.2     \\
$^{16}$C\{$2^+$\}+n        &        -  &       103 &        -  &     9788  &     344   &   37900    &      1  &     10:13     & 0.2     \\
$^{16}$C\{$0^+,2^+$\}+n    &      478  &       144 &  2886779  &    41795  &     473   &  206048    &     15  &     03:59     & 1.4     \\
$^{16}$C\{$0^+,2^+,4^+$\}+n&     1269  &       244 & 40036229  &   118450  &     546   & 2991231    &     16  &     56:01     & 2.5     \\ \hline
\end{tabular}
\caption{\label{TABLE:size} Size of calculations for the different model spaces. The first column gives the projectile model, with the core states included. The next 5 columns give the number of
permutations for various variables that are required. The last 4 columns give total CPU time, the number of CPUs,
walltime, and memory requirements to store the form factors.}
\end{table*}

\begin{table*}\begin{tabular}{|c|c|cccccc|}\hline
type        & interaction                 & $V$       & $r_V$     & $a_V$     & $W$       & $r_W$     & $a_W$     \\ \hline
central     & core + target               & 123.0     & 0.75      & 0.8       & 65.0      & 0.78      & 0.8       \\
deformed    & core + target               & 134.1     & 0.75      & 0.8       & 68.15     & 0.75      & 0.8       \\
central     & neutron + target (E=62 MeV) & 34.54     & 1.17      & 0.75      & 13.4      & 1.26      & 0.58      \\
central     & neutron + target (E=60 MeV) & 33.79     & 1.17      & 0.75      & 12.08     & 1.26      & 0.58      \\ \hline
\end{tabular}
\caption{\label{TABLE:pot-par} Fragment-Target optical potential parameters.
The radius parameters are to be multiplied by $A_\mathrm{core}^{1/3}+A_\mathrm{target}^{1/3}$ for the core-target potential and $A_\mathrm{target}^{1/3}$ for the neutron-target potential.
The core-target parameters are used for both $^{16}$C and $^{10}$Be, and the neutron-target parameters are only adjusted slightly for the different energy of the reaction.}
\end{table*}

\section{\label{results} Results}

Here we present the results of the calculations. For both reactions, we perform three different
calculations to understand the relevance of the different ingredients.
First we simulate the CDCC version of the single particle calculations in \cite{aumann00,madda01}.
We refer to it as {\bf SPIS}: single particle incoherent sum.
It corresponds to a sum over the CDCC single particle cross sections multiplied by the corresponding
relative probability for the projectile generated in Section \ref{structure}. It does not contain
any interference between core states nor core excitation/de-excitation due to the interaction
with the target.
Secondly, we  perform a calculation where we do not include dynamic effects
with the target, and refer to it as {\bf CCSE}: coupled channels static excitation.
It contains the full coupling within the projectile but has only $Q=0$ in the
three-body coupling matrix elements.
The third case is the full calculation, which we denote by {\bf CCDE} for
coupled channels dynamic excitation.
In all the above, we perform the calculations switching on and off continuum-continuum
coupling to determine the importance of these couplings as done before \cite{nunes99}.
In Tables~\ref{TABLE:c17-2+}--\ref{TABLE:be11-n}, the couplings labelled c-v and c-T refer
to the deformed couplings between the core-valence and core-target subsystems.
The monopole couplings are always included.

\subsection{\label{results:c17} $^9$Be($^{17}$C,$^{16}$C+$n$) \@ 62 MeV/nucleon}

The 1-step (no continuum-continuum coupling) breakup cross sections for
$^9$Be($^{17}$C,$^{16}$C$\{0^+,2^+\}$+n) are given in Table~\ref{TABLE:c17-2+}.
The single particle cross sections for each of the configurations used in the CC calculations
are 33 mb, 25 mb and 13 mb for the $[d\otimes0^+]$, $[s\otimes2^+]$, and $[d\otimes2^+]$ states
respectively (we neglect the $[g\otimes2^+]$ state since its contribution is very small).
These are then multiplied by 0.175, 0.103, and 0.721 to obtain the single particle
incoherent sum SPIS cross section of 18 mb.
This model is equivalent to that used widely to analyze knockout data to
specific states \cite{knockout},
the difference being that this is a 1-step breakup reaction model which
includes nuclear and Coulomb breakup whereas the eikonal model includes 
single-particle-continuum-continuum couplings
but nuclear breakup only.
We will include continuum-continuum couplings in our model later.
Introducing a coupled channels description of the projectile (CCSE)
provides very similar cross sections when comparing to the SPIS model.
This suggests little interference between configurations during the reaction.
Including dynamical excitation in the reaction model (CCDE) enhances the cross section to
both core states, increasing the total breakup cross section to 21 mb
in agreement with the findings of Batham {\it et al.} \cite{batham05}.
However in that work, only cross sections summed over all core states were
given.

\begin{table}\begin{tabular}{|c||c|c||c|c||c|}\hline
      & \multicolumn{2}{c||}{Couplings}&               &               &          \\ \cline{2-3}
Model & c-v            & c-T           & $\sigma_{0^+}$& $\sigma_{2^+}$& $\sigma$ \\ \hline
SPIS  & \no            & \no           &    6          &   12          &   18     \\
CCSE  & \yes           & \no           &    5          &   11          &   17     \\
CCDE  & \yes           & \yes          &    7          &   14          &   21     \\ \hline
\end{tabular}
\caption{\label{TABLE:c17-2+} $^{17}$C 1-step (no CC couplings) breakup cross 
sections leading to  $^{16}$C$\{0^+,2^+\}$ + n: SPIS is the single particle incoherent sum model,
where neither core-valence or core-target excitation couplings are included,
CCSE is the coupled channel static excitation calculations, where 
core-valence couplings are included but no core-target, and finally the
full calculation CCDE coupled channel dynamic excitation,
where both core-valence or core-target excitation couplings are included.}
\end{table}

We explore the relevance of other states in $^{16}$C by first calculating the
inelastic reaction $^{9}$Be($^{16}$C$(2^+)$,$^{16}$C$^*$). We use a simple rotational
model for $^{16}$C, and include three states $\{0^+,2^+,4^+\}$.
We have $^{16}$C$(2^+)$ in the initial state since $^{17}$C has
predominantly $^{16}$C$(2^+)$ in the ground state.
The resulting inelastic cross section to specific core states is
$\sigma_{0^+}=4$ mb, $\sigma_{2^+}=5$ mb, and $\sigma_{4^+}=12$ mb.
The large number of projections of the $4^+$ state and
its relatively low energy $E_{4^+}=4.142$ Mev, raises suspicion that this $4^+$ state will hold a
significant contribution to the breakup cross section of $^{17}$C.
We recognize that the $2^+$ excited state in $^{16}$C is not described well
in a collective model, but here we use it as a toy model to compare with previous work.
One may also wonder about contributions from other states in a rotational band, since we
are using the particle-rotor model.
We include a $4^+$ state here since we start with $^{16}$C in a $2^+$ state.
Inelastic excitations with $\Delta I=4$, within our rotational model,
are an order of magnitude smaller than the $\Delta I=2$ transitions,
thus it should only be necessary to include the $\Delta I=2$ states for the core
in our rotational model.

We therefore recalculate the cross section for $^{17}$C breakup including
the $4^+$ state in the continuum (Table~\ref{TABLE:c17-4+}).
We keep the same ground state with only $\{0^+,2^+\}$ so that
the relative weights of each ground state configuration are preserved.
Since we have no $4^+$ in the ground state, the SPIS model is the same as before.
The $4^+$ state enhances the total breakup cross section in the CCSE model from 17 mb to 22 mb and in the CCDE model from 21 to 27 mb.
The last two rows in Table~\ref{TABLE:c17-4+} are calculations including continuum-continuum couplings.
These couplings suppress the cross section as was
seen before in the single particle case \cite{nunes98,summers04}.
It is a mere coincidence, though, that the final total cross section of $19$ mb compares
so well with the SPIS result.

\begin{table}\begin{tabular}{|c||c|c|c||c|c|c||c|}\hline
      & \multicolumn{3}{c||}{Couplings}&         &               &               &          \\ \cline{2-4}
Model & c-v    & c-T   & CC      & $\sigma_{0^+}$& $\sigma_{2^+}$& $\sigma_{4^+}$& $\sigma$ \\ \hline
SPIS  & \no    & \no   & \no     &    6          &   12          &    -          &   18     \\
CCSE  & \yes   & \no   & \no     &    6          &   12          &    5          &   22     \\
CCDE  & \yes   & \yes  & \no     &    7          &   15          &    5          &   27     \\ \hline
SPIS  & \no    & \no   & \yes    &    5          &   11          &    -          &   16     \\
CCDE  & \yes   & \yes  & \yes    &    4          &   10          &    5          &   19     \\ \hline
\end{tabular}
\caption{\label{TABLE:c17-4+} $^{17}$C breakup cross sections leading to $^{16}$C$\{0^+,2^+,4^+\}$ + n.
The abreviations in the first row are the same as in Table \ref{TABLE:c17-2+}. Calculations are
repeated with and without continuum-continuum couplings (labeled here by CC).}
\end{table}

\subsection{\label{results:be11} $^9$Be($^{11}$Be,$^{11}$Be+$n$) \@ 60 MeV/nucleon}

The breakup cross sections for the reaction $^9$Be($^{11}$Be,$^{10}$Be$\{0^+,2^+\}$+n)
are shown in Table~\ref{TABLE:be11}.
The single particle cross sections for 1-step breakup are 159 mb for the $[s\otimes0^+]$
state, and 10 mb for the $[d\otimes2^+]$ state.
Multiplying these single particle cross sections by the relative probabilities of 0.883 and
0.117 respectively,
one gets a total breakup cross section of 141 mb for the SPIS model.
If we calculate the cross section in the CCSE, the cross section drops to 135 mb,
suggesting some destructive interference between the configurations.
When we include dynamical excitation (CCDE) the cross section is enhanced at 150 mb.
The apparent agreement between the cross section to the $0^+$ state (142 mb) in the CCDE model
and the single particle cross section multiplied by the relative probability (140 mb)
is merely a coincidence. It is a result of cancellations
between two different effects, the interference between different configurations of $^{11}$Be and
the dynamical excitation couplings in the core-target potential.
We also see that the SPIS model severely underpredicts the amount of cross section to the $2^+$ state.
The additional cross section seems to come from including the couplings between the $^{10}$Be and the neutron
rather than from the dynamical excitation couplings of the core-target interaction.
When continuum-continuum couplings are included there is a general reduction of
the cross sections, but the same qualitative effects of projectile couplings and
dynamical excitation remain.

\begin{table}\begin{tabular}{|c||c|c|c||c|c||c|}\hline
      & \multicolumn{3}{c||}{Couplings}&         &               &          \\ \cline{2-4}
Model & c-v    & c-T   & CC      & $\sigma_{0^+}$& $\sigma_{2^+}$& $\sigma$ \\ \hline
SPIS  & \no    & \no   & \no     &  140          &    1          &  141     \\
CCSE  & \yes   & \no   & \no     &  128          &    7          &  135     \\
CCDE  & \yes   & \yes  & \no     &  142          &    8          &  150     \\ \hline
SPIS  & \no    & \no   & \yes    &  109          &    1          &  110     \\
CCSE  & \yes   & \no   & \yes    &  107          &    8          &  115     \\
CCDE  & \yes   & \yes  & \yes    &  109          &    8          &  117     \\ \hline
\end{tabular}
\caption{\label{TABLE:be11} $^{11}$Be nuclear and Coulomb breakup cross sections for $^{10}$Be$\{0^+,2^+\}$ + n
The abreviations in the first row are the same as in Table \ref{TABLE:c17-2+}.}
\end{table}

In the eikonal model of breakup reactions used in the analysis of knockout
data \cite{knockout}, only nuclear breakup
is included. In order to compare with previous calculations for this reaction \cite{aumann00},
where the contributions from Coulomb breakup and dynamical excitation were roughly estimated,
we rerun our calculations with only nuclear breakup.
Result for the CCDE show that the nuclear cross section to the $0^+$ state
is 10 mb less than the cross section calculated when both nuclear and Coulomb are included.
This appears to agree well with the estimate of 10 mb given in \cite{aumann00},
however it does not mean that Coulomb and nuclear should be added incoherently.
A Coulomb breakup only calculation yields 12 mb.
The estimated contribution from dynamical excitation of $0^+$ core to the $2^+$ excited state
was 11 mb multiplied by the spectroscopic factor of the $0^+$ core, equivalent to 0.883 in our rotational model, which gives 10 mb.
We see from Table~\ref{TABLE:be11} that the increase in the cross section to
the $2^+$ state, from the SPIS to our CCDE model is 7 mb. So it turns out that the estimates
presented in \cite{aumann00} produced total cross sections close to those we obtain
in a more accurate calculation. Again, it is important to note that in general
this will not happen.
A comparison of the knockout data from \cite{aumann00} within this new model which incorporates core deformation in a consistent manner is given in \cite{summers06}.

\begin{table}\begin{tabular}{|c||c|c|c||c|c||c|}\hline
                   & \multicolumn{3}{c||}{Couplings}&         &               &          \\ \cline{2-4}
Model              & c-v    & c-T   & CC      & $\sigma_{0^+}$& $\sigma_{2^+}$& $\sigma$ \\ \hline
SPIS               & \no    & \no   & \yes    &  103          &    1          &  104     \\
CCDE               & \yes   & \yes  & \yes    &  100          &    7          &  107     \\ \hline
\end{tabular}
\caption{\label{TABLE:be11-n} $^{11}$Be nuclear breakup cross sections for $^{10}$Be$\{0^+,2^+\}$ + n. The abreviations in the first row are the same as in Table \ref{TABLE:c17-2+}.}
\end{table}

\section{\label{conclusions} Summary and Conclusions}

In summary, we have extended the CDCC method to include a coupled channels description of the projectile,
using a deformed core-valence interaction, thus allowing excited core contributions in the projectile states.
We also include a deformed core-target interaction, allowing the core to (de-)excite during the reaction.

For the reactions considered here, the breakup of $^{17}$C and $^{11}$Be on a $^9$Be target at
$\approx$ 60 MeV/nucleon, the comparison of the structure calculations with the experimental
data will depend on the reaction model used.
By including more of the important dynamics of the reaction mechanisms,
a more direct comparison of structure and experiment will be possible.
Even though the excited states of $^{16}$C are not well described by collective excitations,
we use a toy model of rotational excitations to the excited $2^+$ and $4^+$ states.
This is to compare with previous calculations.
In the breakup of $^{17}$C we find that total cross section is significantly increased due to
excitation of the core, which is consistent with previous calculations.
Our calculations here go beyond previous work allowing for a partial wave decomposition of the
final core states.
Even when there is no $4^+$ core-state in the ground state of $^{17}$C,
significant population of the $4^+$ is predicted in the final breakup states.
Continuum-continuum couplings decrease the cross section.

The particle-rotor model for $^{11}$Be describes the
nucleus well, and there is good agreement with the
spectroscopic factors obtained from shell model calculations.
When we allow the $^{10}$Be to deform, we see an increase in the total breakup cross section
compared to the weighted incoherent sum of the single particle cross sections.
The cross section to the $2^+$ state in $^{10}$Be is significantly increased when the core is deformed.

From our calculations we conclude that both the deformed couplings between
the core and valence particles and the deformed couplings between the core and
target are important and should be included in the reaction model.
This eXtended CDCC (XCDCC) reaction model can now be used to describe breakup
reactions where significant contributions from excited core states are needed
to describe the projectile at any point during the reaction process.
The usefulness of XCDCC is varied. 
Near future applications of XCDCC include
$^{11}$Be(p,p$'$)$^{11}$Be*, $^{12}$C($^{11}$Be,$^{10}$Be+n),
and the various modes of $^{8}$B breakup.
A comparison with breakup data to specific core states
will allow accurate checks of the validity of the structure model for the projectile
and thereby improve the comparison between structure calculations and experiment.
We will also be able to look at detailed breakup spectra and analyze resonances
built on excited core states, which could not be studied using previous
reaction models.

The generality of the formalism for including core degrees of freedom in XCDCC
shows that one can go beyond the simple particle-rotor model for the projectile.
Given the radial overlap wave functions between the projectile and the core states,
and the core elastic and inelastic transitions for the core-target scattering,
a more microscopic picture of the reaction can be obtained, including 
important dynamical effects which can have a significant effect on the reaction observables.

\section*{Acknowledgements}

We thank Jeff Tostevin for useful discussions.
We thank the high performance computing center (HPCC) at MSU for
the use of their facilities.
This work is supported by NSCL, Michigan State University,
the U.K. EPSRC through grant GR/T28577,
and the National Science Foundation through grant PHY-0456656.\\

\appendix

\section{Details on the three-body coupling potentials}
\label{details}

Here we give the details of the derivation for  \eq{vct-coupling}.
The expansion in $Q$, \eq{expansionQ}, and $\lambda$, \eq{expansionlambda}, gives us three spherical harmonics which depend on
the $\vec{\hat{R}}$, $\vec{\hat{r}}$ and $\xi$ degrees of freedom.
The expansion in $K$, \eq{expansionK}, gives us two spherical harmonics in $\vec{\hat{R}}$ and $\vec{\hat{r}}$,
 which combine with the spherical harmonics from the $Q$ expansion to give
 two new multipole orders, $\Lambda$ and $\Lambda'$, which relate to the projectile-target
 and core-valence relative motion respectively:
\be\begin{split}
C_{Kk}(\vec{\hat{R}}) C_{\lambda \mu}(\vec{\hat{R}}) =&
\sum_{\Lambda\omega} C_{\Lambda\omega}(\vec{\hat{R}})
\hat{\Lambda}^2 (-1)^\omega \\ &\times
\threej{K}{\lambda}{\Lambda}{k}{\mu}{-\omega}
\threejz{K}{\lambda}{\Lambda} \ ,
\end{split}\ee

\be\begin{split}
C_{K-k}(\vec{\hat{r}}) C_{Q-\lambda,q-\mu}(\vec{\hat{r}}) =
\sum_{\Lambda'\omega'} C_{\Lambda'\omega'}(\vec{\hat{r}})
\hat{\Lambda'}^2 (-1)^{\omega'} \\ \times
\threej{K}{Q-\lambda}{\Lambda'}{-k}{q-\mu}{-\omega'}
\threejz{K}{Q-\lambda}{\Lambda'} \ .
\end{split}\ee

We then combine the three spherical harmonics, for the three coordinates, to
form a tripolar spherical harmonic with zero total angular momentum, which is a
tensor of rank zero, since the matrix elements of the potential operator must
conserve total angular momentum:
\be\begin{split}
C_{\Lambda\omega}(\vec{\hat{R}}) C_{\Lambda'\omega'}(\vec{\hat{r}})
C_{Q-q}(\xi) = \frac{(-1)^{\Lambda'-Q}}{\hat\Lambda}
\threej{\Lambda}{\Lambda'}{Q}{\omega}{\omega'}{-q} \\ \times \tripolar \ .
\end{split}
\label{a3}
\ee
Using these definitions,  one can write the potential operator 
as \eq{vct-coupling}.

In order to calculate the matrix elements, we use the projectile coupling order
defined in Eqs.~(\ref{coup-alpha})\&(\ref{coup-jp}).
The matrix elements of the tripolar spherical harmonic of rank zero is:
\bwt
\be\begin{split}
&\bra{(L[(ls)j,I;J_P])J,J_t;J_T}\tripolar\ket{(L'[(l's)j',I';J_P'])J',J_t';J_T}
\\&=
\delta_{J_t,J_t'}\delta_{J,J'}\delta_{s,s'}
(-1)^{J_P+J+\Lambda+j'+l+l'+s}
\hat{J_P}\hat{J_P'}\hat{L}\hat{L'}\hat{j}\hat{j'}\hat{l}\hat{l'}
\hat{\Lambda}
\\ &~\times
\threejz{\Lambda}{L}{L'}\threejz{\Lambda'}{l}{l'}
\sixj{J_P}{J_P'}{\Lambda}{L'}{L}{J}\sixj{j}{j'}{\Lambda'}{l'}{l}{s}
\ninej{J_P}{J_P'}{\Lambda}{j}{j'}{\Lambda'}{I}{I'}{Q}
\reduced{I}{C_{Q}(\xi)}{I'} \ .
\end{split}\ee
\ewt
By rearranging the summations, one can write the transition
potentials as defined in Eqs.~(\ref{vjt}--\ref{pcoup}).

The reduced matrix element for the core degree of freedom is defined using the
Bohr \& Mottelson definition:
\be
\reduced{I}{C_{Q}(\xi)}{I'} =
\hat{I} \matrixel{I}{C_{Q}(\xi)}{I'}
/ \cg{I'}{K_c'}{Q}{q}{I}{K_c} \ .
\ee
We have left this matrix element separate since the derivation of the coupling
potential is independent of the choice for the core inelastic excitation mode.
The specific form for the matrix element within a rotational model is
presented in Appendix C.

\section{$Q=0$ limit}
\label{qzero}

If one sets $Q=0$ in the transition potential defined in
Eqs.~(\ref{vjt}--\ref{pcoup}), this corresponds to the limit that the core
cannot dynamically excite during the reaction, hence the core spin in the
initial state is the same as the core spin in the final state. In the $Q=0$
limit, the coupling potentials are the same as for single particle breakup as
defined in \cite{fresco}, since $\Lambda=K$,
\be\begin{split}\label{coup-q0}
U^{J_T}_{\alpha:\alpha'}(R) =& \hat{L}\hat{L'}\hat{J_P}\hat{J_P'} (-1)^{J_P+J}
\sum_{K} (-1)^{K} \hat{K}^2 \\ &\times \threejz{K}{L}{L'}
\sixj{J_P}{J_P'}{K}{L'}{L}{J} F^{K}_{J_Pin:J_P'i'n'}(R) \ ,
\end{split}\ee
where the form-factors are the sum over the coupled channels states of the
radial integrals and couplings,
\be
F^{K}_{J_Pin:J_P'i'n'}(R) = \sum_{aa'}
R^{K00}_{ain:a'i'n'}(R) P^{K00:K}_{a:a'} \ .
\ee
The radial integrals are
\be
R^{K00}_{ain:a'i'n'}(R) = \hat{K} \int_0^{R_m}
u^{i*}_{a:n}(r) V^{0K}_{ct}(r,R) u^{i'}_{a':n'}(r) \dif r \ ,
\ee
and the couplings are
\be\begin{split}
P^{K00:K}_{a:a'} =& \delta_{II'} (-1)^{j'+l+l'+s+J_P'+j+I}
\hat{K}\hat{j}\hat{j'}\hat{l}\hat{l'} \\ &\times
\threejz{K}{l}{l'}
\sixj{J_P}{J_P'}{K}{j'}{j}{I}
\sixj{j}{j'}{K}{l'}{l}{s} \ .
\end{split}\ee
One can see from this limit that once the form-factors are calculated, the coupling
potentials have same form as with core excitation.
We associate this limit with a reaction model where the core cannot dynamically
excite due to the interaction with the target.
This limit may still include couplings between the core and valence particles which
can excite/de-excite the core.

\section{Rotational model for the core}
\label{rotor}

In the formalism for XCDCC, we did not assume a model for the core-valence coupling.
Here we evaluate the XCDCC equations using a rotational model for the core.
For the radial dependence, we use a deformed Woods-Saxon potential
\be\label{ws-pot}
V_{vc}(\vec{r},\xi) = \frac{V}{1+\expl{(r-R(\xi))/a_V}} \ ,
\ee
where the radius depends on the quadrupole deformation:
\be
R(\xi) = R_V (1+\beta_2 Y_{20}(\xi)) \ .
\ee
The core-valence potential are numerically calculated by
expanding in multipoles, as defined in \cite{fresco},
\be
V_{vc}(\vec{r},\xi) = \sum_Q V_{vc}^Q(\vec{r}) P_Q(\xi) \ .
\ee
In the rotational model, the eigenstates of the core, $\varphi_I(\xi)$, are proportional to the rotational matrices,
\be
\braket{\xi}{I} = \frac{I}{\sqrt{8\pi^2}} \mathcal{D}^{I}_{K_c0}(\xi) \ .
\ee
The matrix element, for the core degrees of freedom, in the coupling potentials is then
\be
\reduced{I}{C_{Q}(\xi)}{I'} = (-1)^{I'+K_c}
\hat{I} \hat{I'}
\threej{Q}{I}{I'}{0}{-K_c}{K_c} \ ,
\ee
where $K_c$ is the projection of the core spin, $I$, on the body-fixed $z$-axis.


\end{document}